\documentclass{ws-procs9x6-cpt22}
\begin{document}

\newcommand{\refeq}[1]{(\ref{#1})}
\def\etal {{\it et al.}}
%any other macros go here 

\title{Constraining GUP Models Using Limits on SME Coefficients}

\author{Andr\'e H. Gomes}

\address{Departamento de Física, Universidade Federal de Ouro Preto,\\
Ouro Preto, Minas Gerais, Brazil}

\begin{abstract}
In this proceedings, I outline recent efforts to constrain models based on generalized uncertainty principles (GUP) using limits on coefficients of the Standard-Model Extension. Two main results are reported: (1) bounds on isotropic GUP models are improved by a factor of $10^{10}$ compared to previous spectroscopic bounds; and (2) anisotropic GUP models are established and also constrained.
\end{abstract}

\bodymatter

\section{Introduction}

Many approaches to quantum gravity contain some notion of a fundamental length scale appearing in the form of generalized uncertainty principles (GUP).\cite{hossen2013} In general, these are written as $\Delta x_i \Delta p_j \ge \tfrac{\hbar}{2} | \langle f_{ij}(\boldsymbol{x},\boldsymbol{p}) \rangle |$  with $f_{ij} \to \delta_{ij}$ when GUP parameters vanish. One way of accessing physics from GUP at the phenomenological level is postulating $[x_i,p_j]=i\hbar f_{ij}(\boldsymbol{x},\boldsymbol{p})$ for the commutator of position and momentum.\cite{maggiore,kmm} Such modifications of quantum mechanics have been widely studied in the past decades for nonrelativistic isotropic models\cite{igup} but remarkably there seems to be no attempt to date at investigating non isotropic ones. In this proceedings I summarize recent efforts on bounding both isotropic and anisotropic GUP models using current limits on Standard-Models Extension (SME) coefficients.\cite{agup}

\section{Framework}

The SME is a common framework for investigating physics beyond the Standard Model, be it Lorentz invariant or not. At the classical level, its Lagrangian is constructed to contain all possible corrections to General Relativity coupled to the Standard Model.\cite{sme,sme-gravity} In this sense, despite being based on conventional commutators, the SME is expected to contain all predictions derived from GUP at the level of effective physics. For the class of nonrelativistic GUP models described next, this is verified by a simple comparison of the GUP Hamiltonian and the nonrelativistic SME.

Two requirements define the class of GUP models we consider: (1) position commutativity, $[x_i,x_j]=0$, hence space homogeneity; and (2) the right-hand side of $[x_i,p_j]=i\hbar\delta_{ij}$ is modified only by terms even in powers of the momentum. Both restrictions should be relaxed on future works, but here simplify the task of constraining GUP models using bounds on SME coefficients. Requirement (1) suggests consideration of the translation invariant flat spacetime limit of the SME, in particular its nonrelativistic Hamiltonian for free fermions.\cite{sme-fermions} The isotropic limit of this Hamiltonian also reveals odd powers on the momentum always appear coupled to spin degrees of freedom; on the other hand, it is currently unknown how spin affects GUP, thus motivating requirement (2).

\section{Constraints on isotropic GUP models}

The general isotropic deformation of the canonical commutation relation is $[x_i,p_j]=i\hbar[f(p^2)\delta_{ij} + g(p^2) p_i p_j]$ with $gp = f \partial_p f/(f - p \partial_p f)$ to ensure position commutativity. Depending on $f$ there is $(\Delta x_i)_\text{min}>0$; if so, $|\boldsymbol{x}\rangle$ cannot be used as basis for physical states.\cite{min-x,kmm} To be general, since in our case the generator of translations $T_i$ satisfies $[x_i,T_i]=i\hbar\delta_{ij}$, we use its eigenvectors $|\boldsymbol{\rho}\rangle$ and find the wave function representation $x_i\psi(\boldsymbol{\rho})=i\hbar\partial_{\rho_i}\psi(\boldsymbol{\rho})$ and $p_i\psi(\boldsymbol{\rho})=p_i(\boldsymbol{\rho})\psi(\boldsymbol{\rho})$, where $p_i=\rho_i f(p(\rho^2))$.

Perturbatively, we write $p_i(\rho^2) = \rho_i(1 + \sum \alpha_{n}\rho^{n})$ ($n$ even $\ge0$) and split the free Hamiltonian into the conventional $\rho^2/2m_\psi$ and a perturbation
\begin{equation}
\delta H_\text{iGUP} = \frac{ 2\alpha_0+\alpha_0^2 }{ 2m_\psi } \rho^2 + \sum_{ n=4,6,\dots }^\infty \sum_{k=0,2,\dots}^{n-2} \frac{ \overline{\alpha}_{n-k-2} \overline{\alpha}_{k} }{ 2m_\psi } \rho^{n},
\end{equation}
with $\overline{\alpha}_0=1+\alpha_0$ and $\overline{\alpha}_n=\alpha_n$ for $n>0$. This is to be compared with the nonrelativistic isotropic limit of the free fermion SME Hamiltonian,\cite{sme-fermions}
\begin{equation}
\delta H_\text{iLV} = - \sum_{n=0,2,\dots}^\infty  \mathring{c}_{n}^\text{NR} \rho^{n},
\end{equation}
where spin-dependent, as well as CPT-odd terms, have been neglected. Matching same powers in $\rho$ shows effective physics from the isotropic GUP model considered here is entirely contained in the SME framework.

To set constraints on GUP parameters $\alpha_n$, we note currently available bounds on SME coefficients $\mathring{c}_{n}^\text{NR}$ are in the electron sector and comes from an observed $1.8$ $\sigma$ difference in the theoretical and experimental values of the Positronium 1S-2S frequency, suggesting experimental reach of about $10^{-5}\text{ GeV}^{-1}$ for $\mathring{c}_{2}^\text{NR}$ and $10^{5}\text{ GeV}^{-3}$ for $\mathring{c}_{4}^\text{NR}$.\cite{kost-arnaldo} Identification of GUP parameters with SME coefficients, in particular $2\alpha_0+\alpha_0^2 = - 2m_\psi \mathring{c}_{2}^\text{NR}$ and $(1+\alpha_0)\alpha_2 = - m_\psi \mathring{c}_{4}^\text{NR}$, then sets $|\alpha_0| \lesssim 10^{-8}$ and $|\alpha_2| \lesssim 10^{2}\text{ GeV}^{-2}$. Parameter $\alpha_0$ relates to species-dependent mass rescaling of $\hbar$,\cite{maggiore2021} and to the best of our knowledge has not been constrained before. Parameter $\alpha_2$ corresponds to the parameter $\beta$ in the GUP literature, often related to $(\Delta x_i)_\text{min}>0$, and although our bound on it is weaker by a factor of $10^3$ compared to the strongest spectroscopic bound reported in the literature,\cite{stetsko} we see next consideration on non isotropic GUP allows for stronger bounds.

\section{Constraints on anisotropic GUP models}

If Lorentz symmetry is not an exact symmetry of nature, the expression for $\delta H_\text{iGUP}$ holds only in preferred inertial reference frames while invalid in others as sequences of boosts generally induce a relative rotation among frames. Thus, if GUP has its origin in high-energy Lorentz-violating phenomena, its isotropic formulation predicts the subset of rotation-invariant effects only. Wealthier set of predictions is then achieved considering non isotropic formulations of GUP. What we propose here is
\begin{equation}
[x_i,p_j] = i\hbar [ f(\boldsymbol{p})\delta_{ij} + g_i(\boldsymbol{p}) p_j],
\quad
[x_i,x_j] = 0
\,\,\,\,
\longleftrightarrow
\,\,\,\,
g_i = \frac{f \partial_{p_i} f}{f - \boldsymbol{p}\cdot{\partial}_{\boldsymbol{p}} f}.
\end{equation}
Although we do not claim this to be the most general anisotropic GUP model with commuting $x_i$, we were not able to find others.

As in the isotropic case, wave function representation based on $\psi(\boldsymbol{\rho})$ is achievable and $\delta H_\text{aGUP}$ is derived expanding $p_i=\rho_i f(\boldsymbol{p}(\boldsymbol{\rho}))$ in a power series. However, $f$ contains couplings of $\boldsymbol{\rho}$ to background fields and deriving an explicit expansion for general $f$ is challenging. For the sake of illustration, we set $f=1+\beta_{ij}p_i p_j$; hence $[x_i,p_j] \approx i\hbar [(1+\beta_{kl}p_k p_l)\delta_{ij} + 2\beta_{ik}p_k p_j]$ and $[x_i,x_j]=\mathcal{O}(\beta^2)$. Thus, expanding $\delta H_\text{aGUP}$ in spherical harmonics,
\begin{equation}
\delta H_\text{aGUP} = \beta_{ij}\rho_i \rho_j \frac{\rho^2}{m_\psi} = \sum_{\ell m} \frac{\beta_{4\ell m}}{m_\psi} Y_{\ell}^m (\hat{\boldsymbol{\rho}}) \rho^4,
\end{equation}
and noticing $\beta_{ij}$ has 6 linearly independent components, we see $\delta H_\text{aGUP}$ matches with the $n=4$ and $\ell=\{0,2\}$ restriction of the spin-independent CPT-even nonrelativistic limit of the free fermion SME Hamiltonian,\cite{sme-fermions}
\begin{equation}
\delta H_\text{LV} = -\sum_{\ell m} c^\text{NR}_{4\ell m} Y_{\ell}^m (\hat{\boldsymbol{\rho}}) \rho^4 = -\sum_{\ell m} \left( \sum_{d=6,8,\dots}^\infty m_\psi^{d-7} c^{(d)}_{4\ell m} \right) Y_{\ell}^m (\hat{\boldsymbol{\rho}}) \rho^4.
\end{equation}
The nonrelativistic coefficients $c^\text{NR}_{4\ell m}$ are combinations of the relativistic $c^{(d)}_{4\ell m}$, as in the last equality above, and while there is no direct bound on the first, there are on the latter. We assume the least suppressed contribution to $c^\text{NR}_{4\ell m}$ comes from $c^{(6)}_{4\ell m}$, whose cartesian components $c^{(6)\mu\nu\rho\sigma}_\text{eff}$ are currently bounded in the Sun-centered frame by searches for annual variations of the 1S-2S hydrogen transition frequency, $c_\text{eff}^{(6)TTTJ} + c_\text{eff}^{(6)TXXJ} + c_\text{eff}^{(6)TYYJ} + c_\text{eff}^{(6)TZZJ} \lesssim 10^{-4}\text{ GeV}^{-2}$ for $J=\{X,Y,Z\}$.\cite{kost-arnaldo}

To constrain $\beta_{ij}$, we identify $c^{(6)\mu\nu\rho\sigma}_\text{eff}$ with $\beta^{(\mu\nu}\eta^{\rho\sigma)}$, where only $\beta^{ij}\neq0$, $\eta^{\mu\nu}=\text{diag}(+,-,-,-)$, and enclosing parentheses mean symmetrization on all indices. In the Sun-centered frame, the above bound translates into
\begin{equation}
\beta^{TJ} \lesssim 10^{-4} \text{ GeV}^{-2},
\end{equation}
placing effects from this particular coefficient in the realm of nuclear physics. In the laboratory frame, the above SME coefficient combination relates to the isotropic combination $\mathring{c}^{(6)}_2=c_\text{eff}^{(6)0011}+c_\text{eff}^{(6)0022}+c_\text{eff}^{(6)0033}$, thus
\begin{equation}
\beta_{11} + \beta_{22} + \beta_{33} \lesssim 10^{-8} \text{ GeV}^{-2}
\end{equation}
since in this frame there is no suppression due to Earth's orbital speed. From a different perspective, assuming Lorentz symmetry invariance, so $\beta_{ij}\to\beta\delta_{ij}$, this constraint reads
\begin{equation}
\beta \ge 10^{-8} \text{ GeV}^{-2},
\end{equation}
representing improvement by a factor of $10^{10}$ over the current strongest spectroscopic bound\cite{stetsko} on the parameter $\beta$ of isotropic GUP.


\begin{thebibliography}{xx}

\bibitem{hossen2013}
  See, e.g., S.\ Hossenfelder, Liv.\ Rev.\ Relativ.\ \textbf{16}, 2 (2013).
  
\bibitem{maggiore}
M.\ Maggiore, Phys.\ Lett.\ B \textbf{319}, 83 (1993).

\bibitem{kmm}
A.\ Kempf, G.\ Mangano, and R.B.\ Mann, Phys.\ Rev.\ D \textbf{52}, 1108 (1995).

\bibitem{igup}
{\it A framework for non\-relativ\-istic iso\-tropic models based on generalized uncertainty principles,} A.H.\ Gomes, arXiv:2202.02044v2.

\bibitem{agup}
\textit{Deformations of the canonical commutation relation and Lorentz symmetry violation,} A.H.\ Gomes, arXiv:2205.02044v2.

\bibitem{sme}
D. Colladay and V.A. Kosteleck\'y, Phys.\ Rev.\ D \textbf{55}, 6760
(1997);  Phys.\ Rev.\ D \textbf{58}, 116002 (1998).

\bibitem{sme-gravity}
V.A.\ Kosteleck\'y, Phys.\ Rev.\ D \textbf{69}, 105009 (2004).

\bibitem{sme-fermions}
V.A.\ Kosteleck\'y and M.\ Mewes, Phys.\ Rev.\ D \textbf{88}, 096006 (2013).

\bibitem{min-x}
K. Abdelkhalek et al., Phys.\ Rev.\ D \textbf{94}, 123505 (2016).

\bibitem{kost-arnaldo}
V.A.\ Kosteleck\'y and A.J.\ Vargas, Phys.\ Rev.\ D \textbf{92}, 056002 (2015)

\bibitem{maggiore2021}
M.\ Fadel and M.\ Maggiore, Phys.\ Rev.\ D \textbf{105}, 106017 (2022).
  
\bibitem{stetsko}
M.M.\ Stetsko, Phys.\ Rev.\ A \textbf{74}, 062105 (2006).

\end{thebibliography}
\end{document}